\documentclass[10pt,a4paper,reqno]{amsart}
\usepackage{amsfonts,amsmath,amsthm}

\author{James T. Ferguson}
\title[Flat Pencils of Symplectic Connections]{Flat Pencils of Symplectic Connections and Hamiltonian Operators of Degree 2}
\date{April 11, 2007}
\address{Department of Mathematics\\ University of Glasgow\\ Glasgow G12 8QW\\ U.K.}
\email{j.ferguson@maths.gla.ac.uk}

\newtheorem{theorem}{Theorem}[section]
\newtheorem{lemma}[theorem]{Lemma}
\newtheorem{corollary}[theorem]{Corollary}
\newtheorem{result}[theorem]{Result}

\newtheorem{definition}[theorem]{Definition}
\newtheorem{proposition}[theorem]{Proposition}
\newtheorem{example}[theorem]{Example}
\theoremstyle{definition}

\newcommand{\utou}[2]{\frac{\partial\tilde{u}^{#1}}{\partial
u^{#2}}}
\newcommand{\uout}[2]{\frac{\partial u^{#1}}{\partial\tilde{u}^{#2}}}
\newcommand{\utouu}[3]{\frac{\partial^2 \tilde{u}^{#1}}{\partial u^{#2}
\partial u^{#3}}}
\newcommand{\uouut}[3]{\frac{\partial^2 u^{#1}}{\partial\tilde{u}^{#2}
\partial\tilde{u}^{#3}}}

\newcommand{\dbyd}[1]{\frac{\partial~}{\partial #1}}
\newcommand{\dfbyd}[2]{\frac{\partial #1}{\partial #2}}

\newcommand{\qnabla}{\widetilde{\nabla}}

\begin{document}

\begin{abstract}
Bi-Hamiltonian structures involving Hamiltonian operators of
degree 2 are studied. Firstly, pairs of degree 2 operators are
considered in terms of an algebra structure on the space of
1-forms, related to so-called Fermionic Novikov algebras. Then,
degree 2 operators are considered as deformations of hydrodynamic
type Poisson brackets.
\end{abstract}

 \maketitle


 \section{Introduction}
 Hamilton's equations for a finite-dimensional system with
 position coordinates $q^i$ and associated momenta $p_i$,
 \begin{eqnarray*}
  \frac{dq^i}{dt} &=& \dfbyd{H}{p_i}\,,\\
  \frac{dp_i}{dt} &=& -\dfbyd{H}{q^i}\,,
 \end{eqnarray*}
 are understood geometrically as describing the flow of a vector
 field $X_H$ which is associated with the Hamiltonian function
 $H(q^1,\dots,q^n,p_1\dots,p_n)$ by the formula $X_H(f)=\{f,H\}$,
 where $\{\cdot,\cdot\}$ is the Poisson bracket:
 \begin{equation}\label{eqn:canpoisson}
 \{f,g\}=\sum_{i=1}^n\left(\dfbyd{f}{q^i}\dfbyd{g}{p_i}
 -\dfbyd{f}{p_i}\dfbyd{g}{q^i}\right)\,.
 \end{equation}

 More generally, one defines a Poisson bracket on an n-dimensional
 manifold $M$ as a map $C^{\infty}(M)\times C^{\infty}(M)\rightarrow
 C^{\infty}(M)$, $(f,g)\mapsto\{f,g\}$, satisfying, for any functions
 $f,g,h$ on $M$:
 \begin{enumerate}
  \item antisymmetry: $\{f,g\}=-\{g,f\}$\,,
  \item linearity: $\{af+bg,h\}=a\{f,h\}+b\{g,h\}$ for any constants
  $a,b$\,,
  \item product rule: $\{fg,h\}=f\{g,h\}+g\{f,h\}$\,,
  \item Jacobi identity:
  $\{\{f,g\},h\}+\{\{g,h\},f\}+\{\{h,f\},g\}=0$\,.
 \end{enumerate}
 The conditions 1-3 identify $\{\cdot,\cdot\}$ as a bivector: a rank
 two, antisymmetric, contravariant tensor field $\omega$ on $M$. It can
 therefore be represented, by introducing coordinates $\{u^i\}$ on $M$, as a matrix of
 coefficients $\omega^{ij}$, giving
 $$\omega =\omega^{ij}\dbyd{u^i}\otimes\dbyd{u^j}
 =\frac{1}{2}\omega^{ij}\dbyd{u^i}\wedge\dbyd{u^j}\,,$$
 and
 \begin{equation}\label{eqn:fdpbop}
 \{f,g\} =\omega^{ij}\dfbyd{f}{u^i}\dfbyd{g}{u^j}\,.
 \end{equation}

 The Jacobi identity places the following constraint on the components of
 $\omega$:
 \begin{equation}\label{eqn:fdjac}
  \omega^{ir}\dfbyd{\omega^{jk}}{u^r}
  +\omega^{jr}\dfbyd{\omega^{ki}}{u^r}
  +\omega^{kr}\dfbyd{\omega^{ij}}{u^r}=0\,.
 \end{equation}

 If the matrix $\omega^{ij}$ is non-degenerate, we may introduce its
 inverse $\omega_{ij}$, satisfying
 $\omega_{ir}\omega^{rj}=\delta^j_i$. The Jacobi identity for
 $\omega^{ij}$ is equivalent to the closedness of $\omega_{ij}$.
 We refer to a closed non-degenerate two-form as a symplectic
 form, and a manifold equipped with one as a symplectic manifold.
 Darboux's theorem asserts that on any $2n$-dimensional symplectic manifold there
 exists a set of local coordinates
 $\{q^1,\dots,q^n,p_1\dots,p_n\}$ in which the
 Poisson bracket takes the form (\ref{eqn:canpoisson}); i.e. the
 components of $\omega^{ij}$, and so those of $\omega_{ij}$, are
 constant.

 One may also introduce Poisson brackets on infinite-dimensional
 manifolds. The loop space of a finite-dimensional manifold $M$,
 $L(M)$, is the space of smooth maps $u:S^1\rightarrow M$. Poisson
 brackets relating Hamiltonians to flows in $L(M)$ will therefore
 act on functionals mapping $L(M)\rightarrow\mathbb{R}$. In
 \cite{dubnov84},\cite{dubnov89} Dubrovin and Novikov studied the so-called
 Poisson brackets of differential-geometric type, which are of the
 form
 \begin{equation}\label{def:pb}
 \{f,g\}= \int\frac{\delta f}{\delta u^i}
 P^{ij}\left(\frac{\delta g}{\delta u^j}\right) dx
 \end{equation}
 where $u^{i}$ are coordinates on the target space $M$, and $x$ is
 the coordinate on $S^1$. $P^{ij}$ is a matrix of differential operators (in
 $\frac{d}{dx}$), with no explicit dependence on $x$, which is
 assumed to be polynomial in the derivatives $u^i_x,u^i_{xx},\dots$.
 If $\{\cdot,\cdot\}$ defines a Poisson bracket on the loop space then
 $P$ is referred to as a Hamiltonian operator.

 There is a grading on such operators, preserved by
 diffeomorphisms of $M$, given by assigning degree 1 to
 $\frac{d}{dx}$, and degree $n$ to the $n^{th}$ $x$-derivative of each
 field $u^i$. An important class is the hydrodynamic type Poisson
 brackets, which are homogeneous of degree 1:
 $$P^{ij}=g^{ij}(u)\frac{d}{dx}+\Gamma^{ij}_k(u)u^k_x\,.$$

 According to the programme set out by Novikov \cite{novikov85},
 differential-geometric type Poisson brackets on $L(M)$ should be studied in
 terms of finite-dimensional differential geometry on the target space $M$.
 When expanded as a polynomial in $\frac{d~}{dx}$ and the field
 derivatives, the coefficients, which are functions of the fields
 $u^i$ alone, can often be naturally related to known objects of
 differential geometry, or else used to define new ones. In the hydrodynamic case,
 for instance, with $g^{ij}$ non-degenerate, $P$ is Hamiltonian if
 and only if $g^{ij}$ is a flat metric on $M$ and
 $\Gamma^k_{ij}=-g_{ir}\Gamma^{rk}_j$ are the Christoffel symbols of
 its Levi-Civita connection.

 In \cite{dubflatpencil} Dubrovin considered the geometry of
 bi-Hamiltonian structures of Hydrodynamic operators, that is
 pairs of such operators compatible in the sense of
 \cite{magribiham}, that every linear combination of them also
 determines a Poisson bracket. In particular, he introduced a
 multiplication of covectors on $M$ and expressed the
 compatibility of the operators in terms of a quadratic relations
 on this algebra.

 This paper is principally concerned with Hamiltonian operators
 which are homogeneous of degree 2. Section \ref{section:H2O}
 presents the differential geometry of such operators, and in
 particular relates the subclass which can be put into a constant
 form by a change of coordinates on $M$ to symplectic connections.
 Section \ref{section:pencils} then considers pairs of operators
 from this subclass, and the algebraic constraints their
 compatibility places upon the associated multiplication. In section
 \ref{section:bhd12} inhomogeneous bi-Hamiltonian structures
 consisting of a degree 1 and a degree 2 operator are studied.

 \section{Hamiltonian Operators of Degree 2}\label{section:H2O}
 We begin with a review of known results on Hamiltonian operators
 of degree 2:
 \begin{equation}\label{def:h20}
  P^{ij}=a^{ij}\left(\frac{d}{dx}\right)^{2}+b^{ij}_{k}u^{k}_{x}\frac{d}{dx}+c^{ij}_{kl}u^{k}_{x}u^{l}_{x}+c^{ij}_{k}u^{k}_{xx},
 \end{equation}
 in which the matrix $a^{ij}$ is assumed to be non-degenerate.
 Such operators have been considered already in, for example, \cite{potemin86},
 \cite{mokhov98}, \cite{doyle93}, \cite{novikov85}, in which the (conditional)
 Darboux theorem has been discussed. In preparation for the
 bi-Hamiltonian theory we present these results without the use of
 special coordinates.

 Under the change of coordinates $\tilde{u}^i=\tilde{u}^i(u^p)$
 the coefficients in $P^{ij}$ transform as
  \begin{eqnarray}
  \tilde{a}^{ij}&=&\utou{i}{p}\utou{j}{q}a^{pq}\nonumber\,,\\
  \tilde{b}^{ij}_{k}&=&\utou{i}{p}\utou{j}{q}\uout{r}{k}b^{pq}_{r}
  -2\utou{i}{p}\utou{s}{q}\utou{j}{r}\uouut{r}{k}{s}a^{pq}\nonumber\,,\\
    \tilde{c}^{ij}_{k}&=&\utou{i}{p}\utou{j}{q}\uout{r}{k}c^{pq}_{r}
  -\utou{i}{p}\utou{s}{q}\utou{j}{r}\uouut{r}{k}{s}a^{pq}\nonumber\,,\\
  \tilde{c}^{ij}_{kl}&=&\utou{i}{p}\utou{j}{q}\uout{r}{k}\uout{s}{l}c^{pq}_{rs}
  +\utou{i}{p}\utou{j}{q}\uouut{r}{k}{l}c^{pq}_{r}\nonumber\\
  && +
  \utou{i}{p}\utouu{j}{q}{s}\uout{r}{(k}\uout{s}{l)}b^{pq}_{r} +
  \utou{i}{p}\frac{\partial^3 \tilde{u}^j}{\partial u^q \partial
  u^r \partial u^s}\uout{r}{k}\uout{s}{l}a^{pq}\nonumber\\&& +
  \utou{i}{p}\utouu{j}{q}{r}\uouut{r}{k}{l}a^{pq}\,,\label{eqn:trrules}
  \end{eqnarray}
 where the brackets denote symmetrisation. So in particular
 $a^{ij}$ transforms as a rank 2 contravariant tensor on the target space and
 $b^{ij}_{k}$ and $c^{ij}_{k}$ are related to Christoffel symbols
 of connections by $b^{ij}_{k}=-2a^{ir}\bar{\Gamma}^{j}_{rk}$ and
 $c^{ij}_{k}=-a^{ir}\Gamma^{j}_{rk}$. Call these connections
 $\bar{\nabla}$ and $\nabla$ respectively.

 The transformation rules for $c^{ij}_{kl}$ are not determined
 uniquely by those for $P$, since (\ref{def:h20}) sees only the
 part symmetric in $k$ and $l$. To fix $c^{ij}_{kl}$, we always
 assume the antisymmetric part is zero. Denote by $a_{ij}$ the inverse
  of $a^{ij}$ defined by $a_{ir}a^{rj}=\delta^{j}_{i}$.

 \bigskip

 The condition that the operation defined in (\ref{def:pb}) is
 skew-symmetric and satisfies the Jacobi identity places
 constraints on the coefficients appearing in (\ref{def:h20}).
 \begin{theorem}
 \label{thm:hamcond}
 The operator $P$ in equation (\ref{def:h20}) defines a Poisson bracket by equation (\ref{def:pb}) if and only if
  \begin{enumerate}
  \renewcommand{\theenumi}{(\Alph{enumi})}
  \renewcommand{\labelenumi}{\theenumi}
  \item $a^{ij}=-a^{ji}$\label{eqn:hamcond:antisym}\,,
  \item $\nabla_{k}a^{ij}=b^{ij}_{k}-2c^{ij}_{k}$\label{eqn:hamcond:deriv}\,,
  \item
  $a^{ir}\left(b^{jk}_{r}-2c^{jk}_{r}\right)=a^{kr}\left(b^{ij}_{r}-2c^{ij}_{r}\right)$
  \label{eqn:hamcond:cyclic}\,,
  \item $\nabla$ is flat (zero torsion, zero curvature)\label{eqn:hamcond:flat}\,,
  \item
  $c^{ij}_{kl}=c^{ij}_{(k,l)}-a_{pr}c^{ri}_{(k}c^{pj}_{l)}$\label{eqn:hamcond:c4reln}\,.
  \end{enumerate}
 \end{theorem}
 \begin{proof}
 \cite{mokhov98} states that, by virtue of being Hamiltonian, the
 operator (\ref{def:h20}) can be put in the form
 \begin{equation}\label{def:shortho}
  P^{ij}=a^{ij}\left(\frac{d~}{dx}\right)^{2}+b^{ij}_{k}u^{k}_{x}\frac{d}{dx}\,,
 \end{equation}
 by a change of coordinates $u^i=u^i({\tilde u})$,
 and that for an operator of this shorter form to be Hamiltonian
 is equivalent to the three conditions
  \begin{enumerate}
  \renewcommand{\theenumi}{(\alph{enumi})}
  \renewcommand{\labelenumi}{\theenumi}
  \item\label{eqn:mokhov:asymm} $a^{ij}=-a^{ji}$\,,
  \item\label{eqn:mokhov:deriv} $a^{ij},_k=b^{ij}_k$\,,
  \item\label{eqn:mokhov:cyclic} $a^{ir}b^{jk}_r=a^{jr}b^{ki}_r$\,.
  \end{enumerate}

 We first assume that $P$ is a Poisson bracket, so there exists
 the special coordinates in which $P$ takes the form
 (\ref{def:shortho}) and \ref{eqn:mokhov:asymm}-\ref{eqn:mokhov:cyclic} hold. By reversing the change of variables as ${\tilde u}^i={\tilde
 u^i}(u)$, conditions \ref{eqn:hamcond:antisym}-\ref{eqn:hamcond:cyclic} of Theorem \ref{thm:hamcond} are Mokhov's
 three conditions converted to tensorial identities. That $\nabla$
 is flat follows from its Christoffel symbols,
 $\Gamma^k_{ij}=-a_{ir}c^{rk}_j$, being zero in the $u$
 coordinates.

 The formula in condition \ref{eqn:hamcond:c4reln} is derived from the transformation rules above. In
 changing from flat coordinates $u^i$ to coordinates ${\tilde
 u}^i$ they give:
 \begin{eqnarray*}
  \tilde{c}^{ij}_{kl}&=&
  \utou{i}{p}\utouu{j}{q}{s}\uout{s}{(k}\uout{s}{l)}b^{pq}_{r} +
  \utou{i}{p}\frac{\partial^3 \tilde{u}^j}{\partial u^q \partial
  u^r \partial u^s}\uout{s}{k}\uout{s}{l}a^{pq}\\&& +
  \utou{i}{p}\utouu{j}{q}{r}\uouut{r}{k}{l}a^{pq}\,,
 \end{eqnarray*}
 and
 \begin{eqnarray*}
 \tilde{c}^{ij}_{k}&=&
  -\utou{i}{p}\utou{s}{q}\utou{j}{r}\uouut{r}{k}{s}a^{pq}\,,\\
  &=& \utou{i}{p}\utouu{j}{q}{r}\uout{r}{k}a^{pq}\,,
 \end{eqnarray*}
 where the last line has used the identity
 $$\utouu{i}{r}{s}\uout{r}{j}\uout{s}{k} +
 \utou{i}{r}\uouut{r}{j}{k}=0\,,$$
 which is a differential consequence of
 $\utou{i}{r}\uout{r}{j}=\delta^i_j\,.$
 \begin{eqnarray*}
 {\tilde c}^{ij}_{k,l} &=& \dfbyd{{\tilde c}^{ij}_k}{{\tilde u}_l}\\
 &=& \utouu{i}{p}{s}\uout{s}{l}\utouu{j}{r}{q}\uout{r}{k}a^{pq}\\
 &&+\utou{i}{p}\frac{\partial^3 {\tilde u}^j}{\partial u^q\partial
 u^r\partial u^s}\uout{r}{k}\uout{s}{l}a^{pq}\\
 &&+\utou{i}{p}\utouu{j}{q}{r}\uouut{r}{k}{l}a^{pq}\\
 &&+\utou{i}{p}\utouu{j}{q}{r}\uout{r}{k}\uout{s}{l}b^{pq}_s\,,
 \end{eqnarray*}
 from which we see
 $${\tilde c}^{ij}_{kl} ={\tilde c}^{ij}_{(k,l)}
 -\utouu{i}{p}{s}\utouu{j}{r}{q}\uout{s}{(l}\uout{r}{k)}a^{pq}\,.$$
 This last term can be seen to be $${\tilde a}_{pr}{\tilde
 c}^{ri}_{(k}{\tilde c}^{pj}_{l)}\,.$$

 Conversely, if \ref{eqn:hamcond:antisym}-\ref{eqn:hamcond:c4reln} hold, the flatness of $\nabla$ asserts the existence
 of coordinates in which $c^{ij}_k=0$, and condition \ref{eqn:hamcond:c4reln} then
 asserts that $c^{ij}_{kl}=0$ in these coordinates.
 \end{proof}

 \label{sec:withlinh20}
 If we take, as a simple case, an operator $P$ as in (\ref{def:h20}) with
 $b^{ij}_k=2c^{ij}_k$ constants, and assume $c^{ij}_{kl}$ to be
 defined by \ref{eqn:hamcond:c4reln}, then $P$ is Hamiltonian if and
 only if $a^{ij}=A^{ij}_ku^k+A^{ij}_0$ where $A^{ij}_k,A^{ij}_0$
 are constants with $A^{ij}_k=c^{ij}_k-c^{ji}_k$,
 $A^{ir}_lc^{jk}_r=A^{jr}_lc^{ik}_r$,
 $A^{ir}_0c^{jk}_r=A^{jr}_0c^{ik}_r$ and
 $c^{ij}_rc^{rk}+c^{ik}_rc^{rj}_l=0$.

 If we take an algebra $\mathcal{A}$ with basis $\{e^1,\dots,e^n\}$,
 $n=\rm{dim}M$, and use $c^{ij}_k$ and $A^{ij}_0$ to define a
 multiplication, $\circ$\,, and skew-symmetric bilinear form,
 $\langle\cdot,\cdot\rangle$, by $e^i\circ e^j=c^{ij}_re^r$ and
 $\langle e^i,e^j\rangle=A^{ij}_0$, then we may rewrite these conditions as
 \begin{eqnarray}
 e^i\circ e^j -e^j\circ e^i &=& A^{ij}_re^r\,,\nonumber\\
 (I\circ J)\circ K &=& -(I\circ K)\circ J\,, \label{eqn:circfna}\\
 \Lambda(I,J,K) &=& \Lambda(J,I,K)\,, \label{eqn:vindberg}\\
 \text{and}\qquad \langle I,J\circ K\rangle &=& \langle J,I\circ K\rangle\,,\nonumber
 \end{eqnarray}
 for all $I,J,K\in\mathcal{A}$, where $\Lambda$ is the associator
 of $\circ$\,: $\Lambda(I,J,K)=(I\circ J)\circ K -I\circ(J\circ K)$.

 Algebras satisfying conditions (\ref{eqn:circfna}) and
 (\ref{eqn:vindberg}) have  appeared before in \cite{xufermion}, in
 the context of linear hydrodynamic Hamiltonian operators taking
 values in a completely odd superspace, where the following
 definition was proposed:

 \begin{definition}\label{def:fna}
 An algebra $(\mathcal{A},\circ)$ satisfying conditions
 (\ref{eqn:circfna}) and (\ref{eqn:vindberg}) is called a
 Fermionic Novikov algebra.
 \end{definition}

 In \cite{baimenghe02} Fermionic Novikov algebras in dimensions
 2-5 were studied, and the listing therein provides a source of
 examples of Hamiltonian operators of degree two.

 \begin{example}
 \begin{eqnarray*}
 P &=& \left(\begin{array}{cccc} 0&0&0&a\\ 0&0&-a&-b-(t-1)u^1\\
 0&a&0&c-u^2\\ -a&b+(t-1)u^1&-c+u^2&0 \end{array}\right)
 \left(\frac{d~}{dx}\right)^2\\
 && + 2\left(\begin{array}{cccc} 0&0&0&0\\
 0&0&0&u^1_x\\ 0&0&-u^1_x&0\\ 0&\tau u^1_x&u^2_x&u^3_x
 \end{array}\right) \left(\frac{d~}{dx}\right)\\
 && +\left(\frac{1}{a}\right)\left(\begin{array}{cccc} 0&0&0&0\\
 0&0&0&0\\ 0&0&0&(u^1_x)^2\\ 0&0&-(u^1_x)^2&0\end{array}\right)
 +\left(\begin{array}{cccc} 0&0&0&0\\
 0&0&0&u^1_x\\ 0&0&-u^1_{xx}&0\\ 0&\tau u^1_{xx}&u^2_{xx}&u^3_{xx}
 \end{array}\right)
 \end{eqnarray*}
 is Hamiltonian for all values of the constants $a,b,c$ and $\tau$ with $a\neq 0$.
 This is the most general Hamiltonian operator associated in the
 manner discussed above  to the algebra designated $(44)_\tau$ in
 \cite{baimenghe02}.
 \end{example}

 Returning to the general Hamiltonian operator (\ref{def:h20}), it
 can be seen from conditions \ref{eqn:hamcond:deriv} and
 \ref{eqn:hamcond:c4reln} in Theorem \ref{thm:hamcond} that
 the coefficients $b^{ij}_k$ and $c^{ij}_{kl}$ in (\ref{def:h20})
 are completely determined by $a^{ij}$ and $c^{ij}_k$. Thus the Hamiltonian
 operator on $L(M)$ is represented uniquely on $M$ by only these
 latter two objects.

 \begin{theorem}\label{thm:adela}
 There is a one-to-one correspondence between Hamiltonian
 operators of the form (\ref{def:h20}) on $L(M)$ and pairs
 $(a,\nabla)$ on $M$ consisting of a non-degenerate bivector $a^{ij}$ and a
 torsion-free connection $\nabla$ satisfying two conditions:
 firstly, that the curvature of $\nabla$ vanishes, and secondly,
 \begin{equation}\label{eqn:bivectcompat}
 a^{ir}\nabla_ra^{jk}=a^{jr}\nabla_ra^{ki}\,.
 \end{equation}
 The Christoffel symbols, $\Gamma_{ij}^k$, of $\nabla$ are related
 to $c^{ij}_k$ by $c^{ij}_k=-a^{ir}\Gamma_{rk}^j$. We then have
 \begin{eqnarray*}
 b^{ij}_k &=& \nabla_ka^{ij}+2c^{ij}_k\,,\\
 c^{ij}_{kl} &=& c^{ij}_{k,l}-a_{pr}c^{ri}_{(k}c^{pj}_{l)}\,.
 \end{eqnarray*}
 \end{theorem}

 With this, we may verify the following facts \cite{potemin86},\cite{mokhov98}:
 \begin{corollary}\label{cor:symbit}
  For P in (\ref{def:h20}) a Hamiltonian operator we have
  \begin{enumerate}
  \renewcommand{\theenumi}{\arabic{enumi}}
  \renewcommand{\labelenumi}{\theenumi.}
  \item $\Gamma$ is the symmetric part of $\bar{\Gamma}$,
  \item Let
  $\bar{T}^{k}_{ij}=\bar{\Gamma}^{k}_{ij}-\bar{\Gamma}^{k}_{ji}$
  be the torsion of $\bar{\nabla}$.  Then
  $\bar{T}_{ijk}=a_{ir}\bar{T}^{r}_{jk}$ is skew symmetric and
  the forms $\bar{T}=\frac{1}{6}\bar{T}_{ijk}du^{i}\wedge
  du^{j}\wedge du^{k}$ and $a=\frac{1}{2}a_{ij}du^{i}\wedge
  du^{j}$ are related by $3\bar{T}=da$.
  \end{enumerate}
 \end{corollary}
 \begin{proof}
 We begin by noting that equation (\ref{eqn:bivectcompat}) is
 equivalent to the condition
 \begin{equation}\label{eqn:2formcompat}
 \nabla_ka_{ij}=\nabla_ia_{jk}
 \end{equation}
 on the two-form $a_{ij}$.

 In terms of covariant Christoffel symbols,
 Theorem \ref{thm:adela} gives
 \begin{equation}\label{eqn:covbfromc}
 \bar{\Gamma}_{ij}^k =
 \frac{1}{2}a^{kr}\nabla_ra_{ij}+\Gamma_{ij}^k\,,
 \end{equation}
 from which it is clear that
 $\bar{\Gamma}_{(ij)}^k=\Gamma_{ij}^k$.

 We therefore also have
 $$\frac{1}{2}\nabla_ka_{ij}=\bar{\Gamma}_{ijk}-\Gamma_{ijk}\,,$$
 where $\bar{\Gamma}_{ijk}=a_{ir}\bar{\Gamma}^{r}_{jk}$ and
 $\Gamma_{ijk}=a_{ir}\Gamma^{r}_{jk}$. Because $\nabla$ is
 torsion-free we have
 \begin{eqnarray*}
 \bar{T}_{ijk} &=& \bar{\Gamma}_{ijk}-\bar{\Gamma}_{ikj}\,,\\
 &=&
 \bar{\Gamma}_{ijk}-\Gamma_{ijk}-\bar{\Gamma}_{ikj}+\Gamma_{ikj}\,,\\
 &=& \frac{1}{2}\nabla_ka_{ij} -\frac{1}{2}\nabla_ja_{ik}\,,\\
 &=& \nabla_ka_{ij}\,,\\
 &=& \nabla_{[k}a_{ij]}\,,\\
 &=& \frac{1}{3}(da)_{ijk}\,.
 \end{eqnarray*}
 \end{proof}

 \begin{lemma}\label{lem:3equiv}
 For a Hamiltonian operator of the form (\ref{def:h20}), the following three statements,
 presented in both covariant and contravariant forms, are equivalent:
  \begin{enumerate}
  \renewcommand{\theenumi}{\arabic{enumi}}
  \renewcommand{\labelenumi}{\theenumi.}
  \item The 2-form $a$ is closed (and so symplectic), or equivalently
  $a^{ij}$ satisfies equation (\ref{eqn:fdjac}) (and so defines a Poisson
  bracket on $M$ by equation (\ref{eqn:fdpbop}));
  \item $\nabla_{k}a^{ij}=0$, i.e. $\nabla_{k}a_{ij}=0$;
  \item $b^{ij}_{k}=2c^{ij}_{k}$, i.e.
  $\Gamma^{k}_{ij}=\bar{\Gamma}^{k}_{ij}$.
  \end{enumerate}
 \end{lemma}
 \begin{proof}
  We see, from the characterisation of Hamiltonian operators given
  in Theorem \ref{thm:adela},
   \begin{eqnarray*}
  a^{ij}\text{ is Poisson} &\Longleftrightarrow&
  a^{ir}a^{jk}_{,r}+a^{jr}a^{ki}_{,r}+a^{kr}a^{ij}_{,r}=0\\
  &\Longleftrightarrow&
  a^{ir}\nabla_ra^{jk}+a^{jr}\nabla_ra^{ki}+a^{kr}\nabla_ra^{ij}=0\\
  &\Longleftrightarrow& 3a^{kr}\nabla_ra^{ij}=0\\
  &\Longleftrightarrow& \nabla_ka^{ij}=0\,,\\
  &\Longleftrightarrow& b^{ij}_k=2c^{ij}_k\,.
  \end{eqnarray*}
 \end{proof}

 Lemma \ref{lem:3equiv} therefore tells us that in the special
 case where the leading coefficient in $P$ is the inverse of a symplectic
 form, the pair $(a,\nabla)$ defining $P$ can be thought of as
 containing the symplectic form $a_{ij}$, and a torsionless
 connection compatible with it (in the sense that $\nabla a=0$);
 that  is, a symplectic connection. More precisely (see e.g.
 \cite{bieliavsky05}):

 \begin{definition}\label{def:sympcon}
 A symplectic connection on a symplectic manifold $(M,\omega)$ is a smooth connection
 $\nabla$ which is torsion-free and compatible with the symplectic form $\omega$, i.e.
 $$\nabla_XY-\nabla_YX-[X,Y]=0$$
 and
 $$\left(\nabla\omega\right)(X,Y,Z)=X(\omega(Y,Z))-\omega(\nabla_XY,Z)-\omega(Y,\nabla_YZ)=0\,,$$
 where $X,Y$ and $Z$ are vector fields on $M$.
 \end{definition}

 In local coordinates $\{x^i\}$, introducing Christoffel symbols
 $\Gamma_{ij}^k$ for $\nabla$ and writing
 $\omega=\frac{1}{2}\omega_{ij}dx^i\wedge dx^j$, the conditions for
 $\nabla$ to be a symplectic connection read
 $\Gamma_{ij}^k=\Gamma_{ji}^k$, as usual, and
 \begin{equation}\label{eqn:symconcompat}
 \nabla_k\omega_{ij} = \frac{\partial\omega_{ij}}{\partial x^r}
 - \Gamma^r_{ki}\omega_{rj} - \Gamma^r_{kj}\omega_{ir} = 0\,.
 \end{equation}

 This definition is analogous to that of the Levi-Civita
 connection of a pseudo-Riemannian metric, however there is an
 important difference in that the Levi-Civita connection is
 uniquely specified by its metric. From the compatibility
 condition (\ref{eqn:symconcompat}) it can be seen that if $\Gamma_{ij}^k$ are the
 Christoffel symbols of a symplectic connection for $\omega$, then
 the connection with Christoffel symbols
 $\tilde{\Gamma}_{ij}^k=\Gamma_{ij}^k+\omega^{kr}S_{rij}$ is a
 symplectic connection if and only if the tensor $S_{ijk}$ is
 completely symmetric. In \cite{gelfandfedosov} a symplectic
 manifold with a specified symplectic connection is called, in light of \cite{fedosov94}, a
 Fedosov manifold. Here we call the pair $(\omega,\nabla)$ of a
 symplectic form and a symplectic connection a Fedosov structure
 on M, and call the structure flat if $\nabla$ is flat.

 In the discussion of Hamiltonian operators it is convenient to
 work with contravariant quantities. We call
 $$\Gamma^{ij}_k=-\omega^{ir}\Gamma^j_{rk}$$
 the contravariant Christoffel symbols of the symplectic
 connection.

 \begin{result}\label{res:compat}
 The compatibility of $\nabla$ and $\omega$ is equivalent to
 $$\frac{\partial\omega^{ij}}{\partial x^k} =
 \Gamma^{ij}_k-\Gamma^{ji}_k\,.$$
 \end{result}

 \begin{result}\label{res:torsion}
 $\nabla$ being torsion-free is equivalent to
 $\omega^{ir}\Gamma^{jk}_r=\omega^{jr}\Gamma^{ik}_r\,.$
 \end{result}

 The curvature of $\nabla$,
 $$R^{k}_{slt}=\partial_{s}\Gamma^{k}_{lt}-\partial_{l}\Gamma^{k}_{st}+\Gamma^{k}_{sr}\Gamma^{r}_{lt}-\Gamma^{k}_{lr}\Gamma^{r}_{st}\,,$$
 can be expressed in terms of contravariant quantities by raising
 indices as
 $$R^{ijk}_l=\omega^{is}\omega^{jt}R^k_{slt}\,.$$
 This gives
 \begin{result}\label{res:curv}
 $$R^{ijk}_l=
 \omega^{ir}\left(\partial_l\Gamma^{jk}_r-\partial_r\Gamma^{jk}_l\right)
 +\Gamma^{ij}_r\Gamma^{rk}_l +\Gamma^{ik}_r\Gamma^{rj}_l\,.$$
 \end{result}

 Having introduced symplectic connections, we are now in a
 position to interpret the following Darboux theorem for
 Hamiltonian operators of degree 2:
 \begin{theorem}\label{thm:H2ODarboux} \cite{potemin86}
 Given a Hamiltonian operator
 $$P^{ij}=a^{ij}\left(\frac{d}{dx}\right)^{2}+b^{ij}_{k}u^{k}_{x}\frac{d}{dx}+c^{ij}_{kl}u^{k}_{x}u^{l}_{x}+c^{ij}_{k}u^{k}_{xx}$$
 where $a^{ij}$ is non-degenerate, then $P$ can be put in the constant form
 $P^{ij}=\omega^{ij}\left(\frac{d}{dx}\right)^2$ (where $\omega$ is a constant matrix)
 by a change of target space coordinates $\{u^i\}$ if and only
 if $a_{ij}$ is closed. The coordinates in which this happens are
 flat coordinates for the connection $\Gamma_{ij}^k=-g_{ir}c^{rk}_j$
 which can be chosen, using a linear substitution, to be canonical
 coordinates for the symplectic form $a_{ij}=\omega_{ij}$.
 \end{theorem}

 In arbitrary coordinates operators satisfying the conditions of
 Theorem \ref{thm:H2ODarboux} have the form
 \begin{equation}\label{def:cpb}
 P^{ij}=\omega^{ij}\left(\frac{d}{dx}\right)^2
 +2\Gamma^{ij}_ku^k_x\frac{d}{dx} +c^{ij}_{kl}u^k_xu^l_x
 +\Gamma^{ij}_ku^k_{xx}
 \end{equation}
 where $\omega^{ij}$ is the inverse of a symplectic form,
 $c^{ij}_{kl}=\Gamma^{ij}_{(k,l)}-\omega_{pr}\Gamma^{ri}_{(k}\Gamma^{pj}_{l)}$,
 and $\Gamma^{ij}_k$ are the contravariant Christoffel symbols of a flat
 symplectic connection compatible with $\omega$. This class of
 operators on $L(M)$ is therefore in one-to-one correspondence
 with flat Fedosov structures on $M$.

 \section{Flat Pencils of Fedosov Structures}\label{section:pencils}
 In this section we consider pairs of Hamiltonian operators of the
 form (\ref{def:cpb}):
 \begin{eqnarray*}
 P_1^{ij}&=&\omega_1^{ij}\left(\frac{d}{dx}\right)^2
 +2{\Gamma_1}^{ij}_ku^k_x\frac{d}{dx} +{c_1}^{ij}_{kl}u^k_xu^l_x
 +{\Gamma_1}^{ij}_ku^k_{xx}\,,\\
 P_2^{ij}&=&\omega_2^{ij}\left(\frac{d}{dx}\right)^2
 +2{\Gamma_2}^{ij}_ku^k_x\frac{d}{dx} +{c_2}^{ij}_{kl}u^k_xu^l_x
 +{\Gamma_2}^{ij}_ku^k_{xx}\,.
 \end{eqnarray*}

 The first fact to establish is that if $P_1$ and $P_2$ are compatible then all elements of the pencil,
 $P_\lambda=P_1+\lambda P_2$, remain in the class (\ref{def:cpb}).

 \begin{theorem}\label{thm:fdbh}
 If $P_1$ and $P_2$ are compatible then $\omega_1^{ij}$ and
 $\omega_2^{ij}$ form a finite-dimensional bi-Hamiltonian
 structure on the target space.
 \end{theorem}
 \begin{proof}
 $P_\lambda$ could have the general form
 $$P_\lambda^{ij}=a_\lambda^{ij}\left(\frac{d}{dx}\right)^{2}
 +{b_\lambda}^{ij}_{k}u^{k}_{x}\frac{d}{dx}
 +{c_\lambda}^{ij}_{kl}u^{k}_{x}u^{l}_{x}
 +{c_\lambda}^{ij}_{k}u^{k}_{xx}\,,$$
 but clearly
 ${b_\lambda}^{ij}_k=2{\Gamma_1}^{ij}_k+2\lambda{\Gamma_2}^{ij}_k$
 and
 ${c_\lambda}^{ij}_k={\Gamma_1}^{ij}_k+\lambda{\Gamma_2}^{ij}_k$,
 so ${b_\lambda}^{ij}_k=2{c_\lambda}^{ij}_k$, and hence, by Lemma
 \ref{lem:3equiv}, $a_\lambda^{ij}$ satisfies the Jacobi identity
 (\ref{eqn:fdjac}) for all $\lambda$.
 \end{proof}

 So we write
 $$P_\lambda^{ij}=\omega_\lambda^{ij}\left(\frac{d}{dx}\right)^2
 +2{\Gamma_\lambda}^{ij}_ku^k_x\frac{d}{dx}
 +{c_\lambda}^{ij}_{kl}u^k_xu^l_x
 +{\Gamma_\lambda}^{ij}_ku^k_{xx}\,.$$

 An immediate corollary of Theorem \ref{thm:fdbh} is that the
 tensor $L^i_j=\omega_1^{ir}\omega_{2rj}$ has vanishing Nijenhuis
 torsion.

 \subsection{Multiplication of covectors}\label{subsec:multiplication}
 As in \cite{dubflatpencil}, we proceed to understand the
 compatibility conditions on $P_1$ and $P_2$ in terms of the
 algebraic properties of a tensorial multiplication of covectors
 on $M$.
 \begin{definition}\label{def:mult}
 Using the tensors
 \begin{eqnarray*}
 \Delta^{sjk} &=&
 \omega_2^{jr}{\Gamma_1}^{sk}_r-\omega_1^{sr}{\Gamma_2}^{jk}_r\,,\\
 \Delta^{jk}_i &=& \omega_{2is}\Delta^{sjk}\,,
 \end{eqnarray*}
 we define a multiplication $\circ$ of covectors on $M$ by
 $$(\alpha\circ\beta)_i=\alpha_j\beta_k\Delta^{jk}_i\,.$$
 \end{definition}

 \begin{theorem}\label{thm:circmult}
 The compatibility of $P_1$ and $P_2$ is equivalent to
 \begin{eqnarray}
 (I,J\circ K)_2 &=& (J,I\circ K)_2\,,\\
 \text{and}\quad (I\circ J)\circ K &=& 0\,,\label{eqn:circfbna}
 \end{eqnarray}
 for all covectors $I,J,K$ on $M$. Here $(\cdot,\cdot)_2$ is the
 skew-symmetric bilinear form on $T^*M$ induced by $\omega_2^{ij}$,
 i.e. $(I,J)_2=I_rJ_s\omega_2^{rs}$. The compatibility also implies
 \begin{equation}
  \nabla^2_l\Delta^{ij}_k = \nabla^2_k\Delta^{ij}_l\,.
 \end{equation}
 \end{theorem}

 Because of Theorem \ref{thm:fdbh}, we phrase the compatibility of
 $P_1$ and $P_2$ in terms of Fedosov structures on $M$, and break
 the above theorem into stages:
 \begin{definition}\label{def:comfed}
 Two flat Fedosov structures $(\omega_1,\nabla^1)$ and
 $(\omega_2,\nabla^2)$, where $\nabla^1$ and $\nabla^2$ have
 contravariant Christoffel symbols ${\Gamma_1}^{ij}_k$ and
 ${\Gamma_2}^{ij}_k$ respectively, are said to be
 \begin{enumerate}
 \renewcommand{\labelenumi}{(\roman{enumi})}
 \item {\em almost compatible} if and only if
 $(\omega_\lambda,\nabla^\lambda)$ is a Fedosov structure for all
 $\lambda$, where the connection $\nabla^\lambda$ is given by
 ${\Gamma_\lambda}^{ij}_k={\Gamma_1}^{ij}_k+\lambda{\Gamma_2}^{ij}_k$.
 \item {\em almost compatible and flat} if and only if they are almost
 compatible, and in addition the curvature of $\nabla^\lambda$
 vanishes for all $\lambda$\,.
 \item {\em compatible} if and only if they are almost compatible and
 flat, and ${c_\lambda}^{ij}_{kl}={\Gamma_\lambda}^{ij}_{(k,l)}
 -\omega_{\lambda pr}{\Gamma_\lambda}^{ri}_{(k}
 {\Gamma_\lambda}^{pj}_{l)}$ satisfies
 ${c_\lambda}^{ij}_{kl}={c_1}^{ij}_{kl}+\lambda{c_2}^{ij}_{kl}$ for all
 $\lambda$.
 \end{enumerate}
 \end{definition}
 The compatibility of two flat Fedosov structures on $M$ is equivalent
 to the compatibility of the associated Poisson brackets on
 $L(M)$.

 We now turn to the two Fedosov strucutres defined by $P_1$ and
 $P_2$, and to the pair $(\omega_\lambda,\nabla^\lambda)$ defined
 by $P_\lambda$. From the linearity of Result \ref{res:compat} in the
 contravariant symbols it can be seen that
 $\omega_\lambda$ is automatically $\nabla^\lambda$-constant, so
 the almost compatibility of $(\omega_1,\nabla^1)$ and
 $(\omega_2,\nabla^2)$ is equivalent to $\nabla^\lambda$ being
 torsion free, i.e. to
 $$\omega_\lambda^{ir}{\Gamma_\lambda}^{jk}_l =
 \omega_\lambda^{jr}{\Gamma_\lambda}^{ik}_l\,.$$
 In flat coordinates for $\nabla^2$, this condition reduces to
 \begin{equation}\label{eqn:ptf}
 \omega_2^{ir}{\Gamma_1}^{jk}_r
 =\omega_2^{jr}{\Gamma_1}^{ik}_r\,.
 \end{equation}
 Note that we already have
 \begin{equation}\label{eqn:1tf}
 \omega_1^{ir}{\Gamma_1}^{jk}_r
 =\omega_1^{jr}{\Gamma_1}^{ik}_r\,.
 \end{equation}

 \begin{lemma}\label{lem:fedacf}
 If $(\omega_1,\nabla^1)$ and $(\omega_2,\nabla^2)$ are almost
 compatible, then the flatness of $\nabla^\lambda$ is equivalent
 to either, and hence both, of
 \begin{eqnarray}
 \partial_l{\Gamma_1}^{jk}_s-\partial_s{\Gamma_1}^{jk}_l &=& 0
  \label{eqn:pzclin}\\
 \text{and}\quad
 {\Gamma_1}^{ij}_{r}{\Gamma_1}^{rk}_{l}+{\Gamma_1}^{ik}_{r}{\Gamma_1}^{rj}_{l}
 &=& 0\label{eqn:pzcquad}
 \end{eqnarray}
 in the flat coordinates for $\nabla^2$.
 \end{lemma}
 \begin{proof}
  The contravariant curvature of $\Gamma_\lambda$ is
 \begin{eqnarray*}
 {R_\lambda}^{ijk}_{l} &=& \omega_\lambda^{ir}\left(\partial_{l}{\Gamma_\lambda}^{jk}_{r}-\partial_{s}{\Gamma_\lambda}^{jk}_{l}\right)+{\Gamma_\lambda}^{ij}_{r}{\Gamma_\lambda}^{rk}_{l}+{\Gamma_\lambda}^{ik}_{r}{\Gamma_\lambda}^{rj}_{l}\\
 &=&{R_1}^{ijk}_l\\
 &&+\lambda\left\{ \omega_2^{is}\left(\partial_l{\Gamma_1}^{jk}_s-\partial_s{\Gamma_1}^{jk}_l\right)
 +\omega_1^{is}\left(\partial_l{\Gamma_2}^{jk}_s-\partial_s{\Gamma_2}^{jk}_l\right)\phantom{\frac{1}{2}}\right.\\
 &&\left.\phantom{+\lambda\frac{1}{2}}+{\Gamma_2}^{ij}_r{\Gamma_1}^{rk}_l
 +{\Gamma_1}^{ij}_r{\Gamma_2}^{rk}_l
 +{\Gamma_1}^{ik}_r{\Gamma_2}^{rj}_l
 +{\Gamma_2}^{ik}_r{\Gamma_1}^{rj}_l \right\}\\
 &&+\lambda^2{R_2}^{ijk}_l\,,
 \end{eqnarray*}
 which in flat coordinates for ${\Gamma_2}^{ij}_k$ reads
 \begin{eqnarray*}
 {R_\lambda}^{ijk}_{l} &=&
 \omega_1^{ir}\left(\partial_l{\Gamma_1}^{jk}_r-\partial_r{\Gamma_1}^{jk}_l\right)
 +{\Gamma_1}^{ij}_r{\Gamma_1}^{rk}_l
 +{\Gamma_1}^{ik}_r{\Gamma_1}^{rj}_l\\
 &&
 +\lambda\omega_2^{is}\left(\partial_l{\Gamma_1}^{jk}_s-\partial_s{\Gamma_1}^{jk}_l\right)\,.
 \end{eqnarray*}
 The vanishing of the order $\lambda$ term is equivalent to
 equation (\ref{eqn:pzclin}), and with this the vanishing of the
 $\lambda$-independent term is equivalent to (\ref{eqn:pzcquad}).
 \end{proof}

 \begin{lemma}\label{lem:c4pencil}
 If $(\omega_1,\nabla^1)$ and $(\omega_2,\nabla^2)$ are almost
 compatible then the condition ${c_\lambda}^{ij}_{kl}={\Gamma_\lambda}^{ij}_{(k,l)}
 -\omega_{\lambda pr}{\Gamma_\lambda}^{ri}_{(k}
 {\Gamma_\lambda}^{pj}_{l)}$ reads, in the flat coordinates for
 $\nabla^2$,
 \begin{equation}\label{eqn:pc4t}
 {\Gamma_1}^{ij}_{r}{\Gamma_1}^{rk}_{l}-{\Gamma_1}^{ik}_{r}{\Gamma_1}^{rj}_{l}=0\,.
 \end{equation}
 \end{lemma}
 \begin{proof}
 For an arbitrary Fedosov structure $(\omega,\nabla)$ the object
 $c^{ij}_{kl}=\Gamma^{ij}_{(k,l)}-\omega_{pr}\Gamma^{ri}_{(k}\Gamma^{pj}_{l)}$
 can be converted into a quadratic expression in contravariant
 quantities as
 \begin{equation}\label{eqn:c4c}
 \omega^{sk}c^{ij}_{kl}=
 \omega^{sk}\Gamma^{ij}_{(k,l)}-\frac{1}{2}\Gamma^{si}_p\Gamma^{pj}_l
 +\frac{1}{2}\Gamma^{pi}_l\Gamma^{sj}_p\,.
 \end{equation}
  This has similarities to the formula for covariant curvature
 obtained in Result \ref{res:curv}; only certain signs have
 changed. Indeed, if we define a quantity $c^j_{rkl}$ by
 $$c^j_{rkl}dx^r =\frac{1}{2}\left(\nabla_{\partial_k}\nabla_{\partial_l}+\nabla_{\partial_l}\nabla_{\partial_k}\right)dx^j\,,$$
 then $c^{ij}_{kl}=\omega^{ir}c^j_{rkl}$.

 We have two ways of expanding
 $\omega_\lambda^{sk}{c_\lambda}^{ij}_{kl}$, corresponding to
 whether we choose first to substitute it into equation
 (\ref{eqn:c4c}), or to expand the pencil quantities.
 We work in flat coordinates for
 $\nabla^2$; in these, ${c_2}^{ij}_{kl}$ also vanishes. First
 expanding the pencil we have
 \begin{eqnarray*}
 \omega_\lambda^{sk}{c_\lambda}^{ij}_{kl} &=&
 \left(\omega_1^{sk}+\lambda\omega_2^{sk}\right){c_1}^{ij}_{kl}\,,\\
 &=& \omega_1^{sk}{c_1}^{ij}_{kl}
  +\lambda\omega_2^{sk}{c_1}^{ij}_{kl}\,,
 \end{eqnarray*}
 whilst (\ref{eqn:c4c}) gives
 \begin{eqnarray*}
 \omega_\lambda^{sk}{c_\lambda}^{ij}_{kl} &=& \omega_\lambda^{sk}{\Gamma_\lambda}^{ij}_{(k,l)}
  -\frac{1}{2}{\Gamma_\lambda}^{si}_p{\Gamma_\lambda}^{pj}_l
  +\frac{1}{2}{\Gamma_\lambda}^{pi}_l{\Gamma_\lambda}^{sj}_p\,,\\
 &=&
 \left(\omega_1^{sk}+\lambda\omega_2^{sk}\right){\Gamma_1}^{ij}_{(k,l)}
  -\frac{1}{2}{\Gamma_1}^{si}_p{\Gamma_1}^{pj}_l
  +\frac{1}{2}{\Gamma_1}^{pi}_l{\Gamma_1}^{sj}_p\,.
 \end{eqnarray*}

 The order $1$ terms merely express equation (\ref{eqn:c4c}) for $P_1$. Equality of the
 order $\lambda$ terms is equivalent to ${\Gamma_1}^{ij}_{(k,l)} = {c_1}^{ij}_{kl}$
 and so to
 \begin{eqnarray*}
 \omega_1^{sk}{\Gamma_1}^{ij}_{(k,l)} &=&
 \omega_1^{sk}{c_1}^{ij}_{kl}\,,\\
 &=& \omega_1^{sk}{\Gamma_1}^{ij}_{(k,l)}-\frac{1}{2}{\Gamma_1}^{si}_p{\Gamma_1}^{pj}_l
 +\frac{1}{2}{\Gamma_1}^{pi}_l{\Gamma_1}^{sj}_p\,.
 \end{eqnarray*}
 \end{proof}

 \begin{proof}[Proof of Theorem \ref{thm:circmult}]
 Using equation (\ref{eqn:ptf}) in Definition \ref{def:mult} it
 can be seen that in the flat coordinates for $\nabla^2$ we have
 $\Delta^{ij}_k={\Gamma_1}^{ij}_k$. Thus we may regard equations
 (\ref{eqn:ptf}),(\ref{eqn:pzclin}),(\ref{eqn:pzcquad}) and
 (\ref{eqn:pc4t}) as identities on $\Delta^{ij}_k$; the result is
 Theorem \ref{thm:circmult}.
 \end{proof}

 The condition imposed by equation (\ref{eqn:pzcquad}) for an
 almost compatible and flat pair of Fedosov structures on the
 mutliplication $\circ$ is $(I\circ J)\circ K=-(I\circ K)\circ J$,
 i.e. the first condition (\ref{eqn:circfna}) satisfied by the
 multiplication of a Fermionic Novikov algebra. In general (\ref{eqn:vindberg}) is not satisfied even for
 compatible Fedosov structures, however we do have, for two flat
 Fedosov structures, $(\omega_1,\nabla^1)$, $(\omega_2,\nabla^2)$, which are almost compatible,
 \begin{eqnarray*}
 \lefteqn{\omega_1^{ir}\nabla^2_r\Delta^{jk}_l
 -\omega_1^{jr}\nabla^2_r\Delta^{ik}_l}\\
 &\qquad=& \Delta^{ij}_r\Delta^{rk}_l
 -\Delta^{ir}_l\Delta^{jk}_r -\Delta^{ji}_r\Delta^{rk}_l
 +\Delta^{jr}_k\Delta^{ik}_r\,.
 \end{eqnarray*}
 So, in particular, if $\Delta^{ij}_k$ is constant in the flat
 coordinates for $\nabla^2$, almost compatible and flat Fedosov
 structures will define a Fermionic Novikov algebra structure on
 the covectors of $M$.

 In \cite{baimenghe02} it emerged that examples of such
 algebras which do not also satisfy the {\lq Bosonic\rq} relation $(I\circ
 J)\circ K=(I\circ K)\circ J$, and hence $(I\circ J)\circ K=0$, are
 relatively rare. $\nabla^2$-constant multiplications arising from
 pairs of Fedosov structures which are almost compatible and flat,
 but not compatible, such as that given in Example
 \ref{ex:acflat} below, are in this class.

 \subsection{The pencil in flat
 coordinates}\label{subsec:flatcoord}
 We now turn our consideration to the form the pencil takes in the
 flat coordinates for $\nabla^2$. From the elements of the proof
 of Theorem \ref{thm:circmult} we have
 \begin{equation}\label{eqn:pbfc}
 P_\lambda^{ij}=
 \left(\omega_1^{ij}+\lambda\omega_2^{ij}\right)\left(\frac{d}{dx}\right)^2
 +2{\Gamma_1}^{ij}_ku^k_x\frac{d}{dx}
 +{\Gamma_1}^{ij}_{k,l}u^k_xu^l_x +{\Gamma_1}^{ij}_ku^k_{xx}\,.
 \end{equation}
 The Jacobi identity for $P_\lambda$ (without assuming $P_1$ and
 $P_2$ are Hamiltonian themselves) is equivalent to the constraints
 \begin{enumerate}
 \renewcommand{\theenumi}{(\roman{enumi})}
 \renewcommand{\labelenumi}{\theenumi}
 \item\label{eqn:pbfc:w2const} $\omega_2^{ij}$ is constant and antisymmetric,
 \item\label{eqn:pbfc:w1anti} $\omega_1^{ij}$ is antisymmetric,
 \item\label{eqn:pbfc:g1tf}
 $\omega_1^{ir}{\Gamma_1}^{jk}_r=\omega_1^{jr}{\Gamma_1}^{ik}_r$,
 \item\label{eqn:pbfc:compat} ${\omega^{ij}_1}_{,k}={\Gamma_1}^{ij}_k-{\Gamma_1}^{ji}_k$,
 \item\label{eqn:pbfc:ptf}
 $\omega_2^{ir}{\Gamma_1}^{jk}_r=\omega_2^{jr}{\Gamma_1}^{ik}_r$,
 \item\label{eqn:pbfc:g1lin} ${\Gamma_1}^{ij}_{k,l}={\Gamma_1}^{ij}_{l,k}$
 \item\label{eqn:pbfc:g1quad} ${\Gamma_1}^{ij}_r{\Gamma_1}^{rk}_l=0$.
 \end{enumerate}

 \begin{proposition}\label{prop:bvector}
 In a fixed coordinate system $\{u^i\}$ (the flat coordinates for
 $\Gamma_2$), given a constant non-degenerate 2-form
 $\omega_2^{ij}$ and a vector field $B=B^r\partial_r$ satisfying
 \begin{equation}\label{eqn:bg1tf}
 \left(\omega_2^{is}B^r_{,s}-\omega_2^{rs}B^i_{,s}\right)\omega_2^{jp}B^k_{,pr}
 =\left(\omega_2^{js}B^r_{,s}-\omega_2^{rs}B^j_{,s}\right)\omega_2^{ip}B^k_{,pr}
 \end{equation}
 and
 \begin{equation}\label{eqn:bzcquad}
 B^j_{,ir}\omega_2^{rs}B^k_{,sl}=0
 \end{equation}
 then the prescription
 \begin{eqnarray*}
 \omega_1^{ij} &=& -(\mathcal{L}_B\omega_2)^{ij}\, =
 \,\omega_2^{ir}B^j_{,r}-\omega_2^{jr}B^i_{,r} \,,\\
 {\Gamma_1}^{ij}_k &=& \omega_2^{ir}B^j_{,rk}
 \end{eqnarray*}
 satisfies the constraints {\em\ref{eqn:pbfc:w2const}-\ref{eqn:pbfc:g1quad}}. Further, all solutions of {\em\ref{eqn:pbfc:w2const}-\ref{eqn:pbfc:g1quad}} have
 this form.
 \end{proposition}

 \begin{proof}
 Equations (\ref{eqn:bg1tf}) and (\ref{eqn:bzcquad}) are the
 quadratic constraints,
 $\omega_1^{ir}{\Gamma_1}^{jk}_r=\omega_1^{jr}{\Gamma_1}^{ik}_r$ and
 ${\Gamma_1}^{ij}_r{\Gamma_1}^{rk}_l=0$ respectively.
 That $\omega_1$ and ${\Gamma_1}$ satisfy the (linear) constraints
 \ref{eqn:pbfc:compat}, \ref{eqn:pbfc:ptf} and \ref{eqn:pbfc:g1lin} is an immediate consequence of their definition.

 Using
 the Poincare lemma together with the symmetries expressed in
 conditions \ref{eqn:pbfc:g1lin} and \ref{eqn:pbfc:ptf}, we have the existence of a vector field
 satisfying ${\Gamma_1}^{ij}_k=\omega_2^{ir}A^j,_{rk}\,$. With this
 condition \ref{eqn:pbfc:compat} gives
 $\omega_1^{ij}=-(\mathcal{L}_A\omega_2)^{ij}+c^{ij}$, where
 $c^{ij}$ is a constant antisymmetric matrix. We may now introduce
 a vector field $B$ with $B^i=A^i+\frac{1}{2}x^sw_{2sr}c^{ri}$
 which satisfies $\omega_1^{ij}=-\mathcal{L}_B\omega_2^{ij}$ and
 ${\Gamma_1}^{ij}_k=\omega_2^{ir}B^j_{,rk}$\,.
 \end{proof}

 Since $\omega_2$ is a symplectic form, its symmetries are
 precisely (locally) Hamiltonian vector fields. Therefore, if
 $\omega_2$ and $\omega_1$ are given, the requirement that
 $\omega_1^{ij}=-\mathcal{L}_B\omega_2^{ij}$ fixes the
 non-Hamiltonian part of $B$. Then the condition
 ${\Gamma_1}^{ij}_k=\omega_2^{ir}B^j_{,rk}$ fixes the Hamiltonian to
 within a quadratic function. From the point of view of the
 multiplication of covectors from Section
 \ref{subsec:multiplication}, the Hamiltonian affects only the
 commutative part of $\circ$, thus the anti-commutative part is
 fixed by $\omega_1^{ij}$ and $\omega_2^{ij}$.

 With consideration of the transformation rules
 (\ref{eqn:trrules}), one can phrase Proposition
 \ref{prop:bvector} as the existence of a vector field $B$ such that
 \begin{eqnarray}
 \omega^{ij}_1 &=& -\mathcal{L}_B\omega^{ij}_2\,,\nonumber\\
 {\Gamma_1}^{ij}_k &=&
 -\mathcal{L}_B{\Gamma_2}^{ij}_k\,.\label{eqn:liederiv1}
 \end{eqnarray}
 We can also calculate from (\ref{eqn:trrules}) the correct
 interpretation of the Lie derivative for an object of type
 $c^{ij}_{kl}$, namely:
 \begin{eqnarray*}
 \mathcal{L}_Xc^{ij}_{kl} &=& X^rc^{ij}_{kl,r} -X^i_{,r}c^{rj}_{kl}
 -X^j_{,r}c^{ir}_{kl} +X^r_{,k}c^{ij}_{rl} +X^r_{,l}c^{ij}_{kr}\\
 &&+X^r_{,kl}c^{ij}_r -\frac{1}{2}X^j_{rl}b^{ir}_k
 -\frac{1}{2}X^j_{,rk}b^{ir}_l -X^j_{,rkl}a^{ir}\,.
 \end{eqnarray*}

 If we work in the flat coordinates for ${\Gamma_2}$, so that the
 components ${c_2}^{ij}_{kl}=0$, we have for our pencil
 \begin{eqnarray*}
 -\mathcal{L}_B{c_2}^{ij}_{kl} &=& +\omega_2^{ir}B^j_{,rkl}\,,\\
 &=& (\omega_2^{ir}B^j_{,rk})_{,l}\,,\\
 &=& {\Gamma_1}^{ij}_{k,l}\,.
 \end{eqnarray*}
 Now, in the flat coordinates for $\nabla^2$ we have the relation
 ${c_1}^{ij}_{kl}={\Gamma_1}^{ij}_{k,l}$. The linearity of the
 transformation rules shows that the Lie derivative of
 ${c_2}^{ij}_{kl}$ should be an object of the same type as
 ${c_1}^{ij}_{kl}$. Thus we have, in addition to
 (\ref{eqn:liederiv1}),
 $${c_1}^{ij}_{kl}=-\mathcal{L}_B{c_2}^{ij}_{kl}\,.$$

 One may understand these three infinitesimal relations between
 the coefficients of $P_1$ and $P_2$ as averring the existence on $L(M)$ of
 an evolutionary vector field
 $$\hat{B}=B^i(u(x))\dbyd{u^i(x)}+\dots$$
 such that
 $$P_1^{ij}=-\mathcal{L}_{\hat{B}}P_2^{ij}\,.$$

 We now turn our attention to some examples of pairs of Fedosov
 structures, using the framework of Proposition
 \ref{prop:bvector}.

 \begin{example}\label{ex:2dma}{\rm Two-dimensional pencils}. Without loss of
 generality we take
 $$\omega_2=\dbyd{u^1}\wedge\dbyd{u^2}\,,$$
 where $u^1$ and $u^2$ are a flat coordinate system for
 $\nabla^2$.

 We take
 $$B=f(u^1,u^2)\dbyd{u^1}+g(u^1,u^2)\dbyd{u^2}$$
 and from it calculate $\omega_1$ and $\Gamma_1$ according to
 (\ref{eqn:liederiv1}). In particular
 $$\omega_1=(f_{,1}+g_{,2})\omega_2\,,$$
 from which it follows immediately that $(\omega_1,\nabla^1)$ and
 $(\omega_2,\nabla^2)$ are almost compatible.

 They are almost compatible and flat if and only if $h=f+\lambda g$ satisfies
 the homogeneous Monge-Ampere Equation $h_{12}^2-h_{11}h_{22}=0$ for all $\lambda$.

 They are compatible if and only if $a=f+\lambda g$ and $b=f+\mu g$ satisfy
 $$a_{12}b_{12}-a_{11}b_{22}=0$$
 for all $\lambda$, $\mu$.

 For instance, one may recover the three two-dimensional Fermionic
 Novikov algebras of \cite{baimenghe02} as constant
 multiplications via
 \begin{enumerate}
  \renewcommand{\theenumi}{(T\arabic{enumi})}
  \renewcommand{\labelenumi}{\theenumi}
  \item\label{T1} $f=u^1$, $g=0$\,,
  \item\label{T2} $f=u^1$, $g=(u^1)^2$\,,
  \item $f=(u^1)^2$, $g=0$\,.
 \end{enumerate}
 \end{example}

 \begin{example}{\rm Commutative algebras.} In the case in which $\omega_1$ is constant
 in the flat coordinates for $\nabla^2$, we have, by condition \ref{eqn:pbfc:compat},
 $${\Gamma_1}^{ij}_k={\Gamma_1}^{ji}_k\,,$$ so that the
 multiplication $\circ$ is commutative.

 In particular if
 $$\omega_1=\omega_2=\omega=\sum_{i=1}^{n}\dbyd{q^i}\wedge\dbyd{p_i}\,,$$
 then the non-Hamiltonian part of $B$ is
 $$\sum_{i=1}^{n}q^i\dbyd{q^i}\,.$$

 To this we may add a Hamiltonian vector field, giving
 $$B=\sum_{i=1}^n\left(\left[q^i+\frac{\partial H}{\partial
 p_i}\right]\dbyd{q^i} -\frac{\partial H}{\partial
 q^i}\dbyd{p_i}\right)\,.$$

 Since $\omega_1=\omega_2$, equation (\ref{eqn:bg1tf}) is immediate.
 Equation (\ref{eqn:bzcquad}) becomes
 $$H,_{ijr}\omega^{rs}H,_{skl}=0\,,$$
 where the indices $i,j,k,l,r,s$ account for both $q$ and $p$
 variables.

 A solution to this is $H=f(x^1,x^2,\dots,x^n)$, where each $x^i$ is either
 $p_i$ or $q^i$; only one from each pair of conjugate variables
 features in $H$.
 \end{example}

 It is not hard to see that Proposition \ref{prop:bvector} can be
 modified to describe almost compatible and flat pairs of Fedosov
 structures. Specifically, we replace equation (\ref{eqn:bzcquad})
 by the expression corresponding to
 ${\Gamma_1}^{ij}_r{\Gamma_1}^{rk}_l={\Gamma_1}^{ik}_r{\Gamma_1}^{rj}_l$,
 namely:
 \begin{equation}\label{eqn:bacflat}
 B^j_{,ir}\omega_2^{rs}B^k_{,sl} =
 B^j_{,lr}\omega_2^{rs}B^k_{,si}\,.
 \end{equation}

 \begin{example}\label{ex:acflat}
 The Fedosov structures specified by
 \begin{eqnarray*}
 \omega_2 &=&
 \dbyd{q_1}\wedge\dbyd{p_1}+\dbyd{q_2}\wedge\dbyd{p_2}\,,\\
 {\Gamma_2}^{ij}_k &=& 0\,,\\
 B &=& \frac{3}{2}q_1^2\dbyd{q_1} +2q_1q_2\dbyd{q_2}
 +q_1p_2\dbyd{p_2}\,,\\
 \end{eqnarray*}
 and $\omega_1^{ij}=-\mathcal{L}_B\omega_2^{ij}$ and
 ${\Gamma_1}^{ij}_k=-\mathcal{L}_B{\Gamma_2}^{ij}_k$
 are almost compatible and flat, but not compatible.

 The non-zero components of $\omega_1$ and
 $\circ$ are
 \begin{eqnarray*}
 \{q_1,p_1\}_1 = \{q_2,p_2\}_1 &=& 3q_1\,,\\
 \{q_2,p_1\}_1 &=& 2q_2\,,\\
 \{p_2,p_1\}_1 &=& p_2\,,
 \end{eqnarray*}
 and
 \begin{eqnarray*}
 dq_2\circ dp_2 &=& dq_1\,,\\
 dp_1\circ dq_1 &=& -3dq_1\,,\\
 dp_1\circ dq_2 &=& -2dq_2\,,\\
 dp_1\circ dp_2 &=& -dp_2\,,\\
 dp_2\circ dq_2 &=& -2dq_1\,.
 \end{eqnarray*}

 Thus, the products
 \begin{eqnarray*}
 (dp_1\circ dq_2)\circ dp_2 &=& -2dq_1\\
 \text{and}\qquad (dp_1\circ dp_2)\circ dq_2 &=& 2dq_1
 \end{eqnarray*}
 violate equation (\ref{eqn:circfbna}) but not
 (\ref{eqn:circfna}). Note that $\circ$ also satisfies
 (\ref{eqn:vindberg}) and thus defines a Fermionic Novikov
 algebra which is not {\lq Bosonic\rq}.
 \end{example}

 \subsection{$\omega N$ manifold with Potential}\label{subsec:wN}
 The tangent bundle $T^*Q$ of a manifold $Q$ is naturally equipped
 with a symplectic form, and thus cotangent bundles form the basic
 set of examples of symplectic manifolds. One may hope to find
 examples of finite-dimensional bi-Hamiltonian structures on
 cotangent bundles by exploiting the existence of additional
 structures on the underlying manifolds. The main object used to
 do this is a $(1,1)$-tensor $L^i_j$ on $Q$ whose Nijenhuis
 torsion is zero. Such an object was utilised by Benenti
 \cite{benenti92} to demonstrate the separability of the geodesic
 equations on a class of Riemannian manifolds. This result was
 later interpreted in \cite{ibortmagrimarmo} in terms of a
 bi-Hamiltonian structure on $T^*Q$ which was extended to a
 degenerate Poisson pencil on $T^*Q\times\mathbb{R}$.

 To obtain Fedosov structures we require more than just a tensor
 $L^i_j$ on $Q$ with vanishing Nijenhuis torsion; we also need a means
 of specifying the connections. If $Q$ is equipped with a
 torsion-free connection $\qnabla$, then the Nijenhuis torsion of
 a $(1,1)$-tensor $L^i_j$ can be written as
 $$N^i_{jk}=L^s_j\qnabla_sL^i_k -L^s_k\qnabla_sL^i_j
 -L^i_s\qnabla_jL^s_k +L^i_s\qnabla_kL^s_j\,.$$
 If there exists a vector field, $A$, on Q such that
 $L^i_j=\qnabla_jA^i$ then
 $$N^i_{jk}=(\qnabla_jA^s)(\qnabla_s\qnabla_kA^i)
 -(\qnabla_kA^s)(\qnabla_s\qnabla_jA^i)
 -(\qnabla_sA^i)(R^s_{jkr}A^r)\,,$$
 where $R^i_{jkl}$ is the curvature tensor of $\qnabla$.

 So, if $\qnabla$ is flat then the vanishing of the Nijenhuis
 tensor of $L=\qnabla A$ is equivalent to the identity
 \begin{equation}\label{eqn:Aznt}
 (\qnabla_jA^s)(\qnabla_s\qnabla_kA^i)
 =(\qnabla_kA^s)(\qnabla_s\qnabla_jA^i)\,.
 \end{equation}

 \begin{proposition}
 Given a manifold $Q$ endowed with a flat connection $\qnabla$ and
 a vector field $A$ satisfying (\ref{eqn:Aznt}), the cotangent
 bundle $T^*Q$ is endowed with a compatible pair of Fedosov
 structures, $(\omega_1,\nabla^1)$ and $(\omega_2,\nabla^2)$, as follows:
 $\omega_2$ is the canonical Poisson bracket on $T^*Q$.

 The connection $\nabla^2$ on $T^*Q$ is the horizontal
 lift \cite{yanokentarobook} of the connection $\qnabla$ on $Q$; i.e. the Christoffel
 symbols ${\Gamma_2}_{ij}^k$ of $\nabla^2$ are zero in the
 coordinates induced on $T^*Q$ by the flat coordinates for
 $\qnabla$.

 $(\omega_1,\nabla^1)$ is calculated from $(\omega_2,\nabla^2)$
 according to the prescription of Proposition \ref{prop:bvector},
 where the vector field $B$ is the horizontal lift of $A$ to
 $T^*Q$.
 \end{proposition}

 \begin{proof}
 Let $\{q^1,\dots,q^n\}$ be flat coordinates for $\qnabla$ on $Q$, and
 $\mathcal{C}=\{q^1,\dots,q^n,p_1,\dots,p_n\}$ be the induced coordinates on $T^*Q$. Then
 $$\omega_2=\sum_{r=1}^n\dbyd{q^r}\wedge\dbyd{p_r}$$
 and
 $$B=\sum_{r=1}^nA^i\dbyd{q^i}\,.$$
 The space of sections of the cotangent bundle of $T^*Q$, $\Omega$, naturally splits into
 $\mathcal{P}={\rm span}\{dp_i\}$ and $\mathcal{Q}={\rm
 span}\{dq^i\}$. For ${\Gamma_1}^{ij}_k=\omega_2^{ir}B^j_{,rk}$ to
 be non-zero requires $k$ to represent a variable $q^k$, and
 $i$ to represent a $p_i$ variable. Thus
 $\Omega\circ\Omega\subseteq\mathcal{Q}$ and
 $\mathcal{Q}\circ\Omega=\{0\}$, meaning that
 $(\Omega\circ\Omega)\circ\Omega=\{0\}$. So the relation
 (\ref{eqn:bzcquad}), ${\Gamma_1}^{ij}_r{\Gamma_1}^{rk}_l=0$, is
 satisfied.

 $\omega_1^{ij}$ has only one kind of non-zero component,
 $\omega^{p_iq^j}=A^j_{,i}$, so the expression
 $\omega_1^{ir}{\Gamma_1}^{jk}_r$ has only one non-zero case:
 $$\sum_{x^r\in\mathcal{C}}\omega_1^{p_ix^r}{\Gamma_1}^{p_jq^k}_{x^r}
 =\sum_{r=1}^n\omega_1^{p_iq^r}{\Gamma_1}^{p_jq^k}_{q^r} =A^r,_iA^k,_{rj}\,,$$
 which is seen to be symmetric in $i$ and $j$ by condition
 (\ref{eqn:Aznt}), which in the flat coordinates $q^i$ reads
 $$A^s,_jA^i,_{sk}=A^s,_kA^i,_{sj}\,.$$
 \end{proof}

 \begin{example}
 If the eigenvalues of $L:TQ\rightarrow TQ$ are functionally
 independent in some neighbourhood then they may be used
 as coordinates, and $L$ takes the form
 $$L=\sum_{i=1}^{n}u^i\dbyd{u^i}\otimes du^i\,.$$

 In this case we may set
 $A=\sum_{i=1}^n\frac{1}{2}(u^i)^2\dbyd{u^i}$, and have $\qnabla$
 defined by vanishing Christoffel symbols in these
 coordinates.

 This gives, writing $v_i$ as the conjugate coordinate to $u^i$ on
 $T^*Q$,
 \begin{eqnarray*}
  \omega_2 &=& \sum_{i=1}^n\dbyd{u^i}\wedge\dbyd{v_i}\,,\\
  \omega_1 &=& \sum_{i=1}^nu^i\dbyd{u^i}\wedge\dbyd{v_i}\,,\\
  {\Gamma_2}^{ij}_{k} &=& 0\\
  {\Gamma_1}^{v_iu^i}_{u^i} &=& -1\,,\\
 \end{eqnarray*}
 and all other Christoffel symbols zero.
 \end{example}

 \section{Bi-Hamiltonian Structures in Degrees 1 and
 2}\label{section:bhd12}
 We now consider a pair of operators, $P_1$ and $P_2$
 in which $P_1$ is a Hamiltonian operator of hydrodynamic type and $P_2$ is of second
 order, i.e.\,:
 \begin{eqnarray*}
 P_1^{ij} &=& g^{ij}(u)\frac{d}{dx}+\Gamma^{ij}_k(u)u^k_x\,,\\
 P_2^{ij} &=& a^{ij}\left(\frac{d}{dx}\right)^{2}+b^{ij}_{k}u^{k}_{x}\frac{d}{dx}+c^{ij}_{kl}u^{k}_{x}u^{l}_{x}+c^{ij}_{k}u^{k}_{xx}\,,
 \end{eqnarray*}
 where $g^{ij}$ is the inverse of a flat metric $g_{ij}$ on $M$
 and $\Gamma^{ij}_k=-g^{ir}\Gamma^j_{rk}$ where the
 $\Gamma^k_{ij}$ are the Christoffel symbols of the Levi-Civita
 connection of $g$. We also assume that $P_2^{ij}$ is
 antisymmetric, so that $a^{ij}=-a^{ji}$,
 $b^{ij}_k=a^{ij}_{,k}+c^{ij}_k+c^{ji}_k$ and
 $c^{(ij)}_{kl}=c^{(ij)}_{(k,l)}$.

 The motivation \cite{dubzhangnorm} for studying such pairs of operators comes not
 from regarding them as separate Hamiltonian operators, but from
 thinking of $P_2^{ij}$ as a first order (dispersive) deformation
 of $P_1^{ij}$ into some non-homogeneous Hamiltonian operator
 $P^{ij}=P_1^{ij}+\varepsilon P_2^{ij}
 +O(\varepsilon^2)$. Thus, in such a pair, it is sensible to
 regard the geometry of $P_1^{ij}$ as being more
 intrinsic than any associated to $P_2^{ij}$.

 We choose to work in flat coordinates for $g$ so that $g^{ij}$ is
 constant and $\Gamma^{ij}_k=0$. Direct calculation of the Jacobi identity for $P^{ij}$ in these
 coordinates yields
 \begin{theorem}\label{thm:d12compat}
 $P_2$ is an infinitesimal deformation of $P_1$, i.e.
 $P^{ij}=P_1^{ij}+\varepsilon P_2^{ij}+O(\varepsilon^2)$ satisfies
 the Jacobi identity to order $\varepsilon$, if and only if
 \begin{enumerate}
 \renewcommand{\theenumi}{(\Roman{enumi})}
 \renewcommand{\labelenumi}{\theenumi}
 \item\label{eqn:d12compat:1} $g^{ir}c^{jk}_r+g^{jr}c^{ik}_r=0$\,,
 \item\label{eqn:d12compat:2} $c^{ij}_{kl}=c^{ij}_{(k,l)}$\,,
 \item\label{eqn:d12compat:3}
 $g^{ir}c^{jk}_{l,r}=g^{jr}(c^{ik}_{l,r}-c^{ik}_{r,l})$\,,
 \item\label{eqn:d12compat:4}
 $g^{ir}(a^{jk}_{,r}-c^{jk}_r)+g^{jr}(a^{ki}_{,r}-c^{ki}_r)+g^{kr}(a^{ij}_{,r}-c^{ij}_r)=0$
 \end{enumerate}
 in the flat coordinates for $g^{ij}$.
 \end{theorem}

 By introducing the tensor
 $T^{ij}_k=a^{ir}\Gamma^j_{rk}+c^{ij}_k$ is it easy to convert
 conditions \ref{eqn:d12compat:1}, \ref{eqn:d12compat:3} and
 \ref{eqn:d12compat:4} to arbitrary coordinates, whilst condition
 \ref{eqn:d12compat:2} becomes
 $$2c^{ij}_{kl}=c^{ij}_{k,l}+c^{ij}_{l,k} -c^{ri}_k\Gamma^j_{rl}
 -c^{ri}_l\Gamma^j_{rk} +T^{ij}_r\Gamma^r_{kl}
 +T^{rj}_k\Gamma^i_{rl} +T^{rj}_l\Gamma^i_{rk}\,.$$

 To consider a bi-Hamiltonian structure involving operators
 $P_1^{ij}$ and $P_2^{ij}$ one need only add conditions
 \ref{eqn:hamcond:cyclic}, \ref{eqn:hamcond:flat} and
 \ref{eqn:hamcond:c4reln} of Theorem \ref{thm:hamcond} to Theorem
 \ref{thm:d12compat}, however, condition \ref{eqn:d12compat:2}
 above allows \ref{eqn:hamcond:c4reln} to be replaced by
 $c^{ij}_rc^{rk}_l=c^{ik}_rc^{rj}_l$.

 \begin{example}
 As discussed in section \ref{sec:withlinh20}, $P_2$ with $b^{ij}_k=2c^{ij}_k$
 constant and $a^{ij}$ non-degenerate is Hamiltonian if and only if
 $a^{ij}=A^{ij}_ku^k+A^{ij}_0$ with $A^{ij}_k=c^{ij}_k=c^{ji}_k$,
 $A^{ij}_0$ is constant, $c^{ij}_k$ are the structure constants of
 a Fermionic Novikov algebra $(\mathcal{A},\circ)$, and $A^{ij}_0$
 defines a skew-symmetric bilinear form on $\mathcal{A}$
 satisfying $\langle I,J\circ K\rangle=\langle J,I\circ K\rangle$.

 If we ask that $P_2$ satisfies the above constancy conditions in
 the flat coordinates for $g^{ij}$, then, defining an inner
 product on $\mathcal{A}$ by $(e^i,e^j)=g^{ij}$, we have that the
 compatibility of $P_1$ and $P_2$ is equivalent to the additional
 constraints:
 \begin{eqnarray*}
 (I\circ J)\circ K &=& (I\circ K)\circ J\,,\\
 (I,J\circ K) &=& -(J,I\circ K)
 \end{eqnarray*}
 and
 $$\left(I,[J,K]\right)+ \left(J,[K,I]\right)
 +\left(K,[I,J]\right)=0\,,$$
 where $[I,J]=I\circ J-J\circ I$ is the commutator of $\circ$,
 which is a Lie bracket by equation (\ref{eqn:vindberg}).

 For example, if we take the algebra $(\mathcal{A}={\rm
 span}\{e^1,e^2,e^3,e^4\},\circ)$ where the only non-zero products
 are $e^3\circ e^3=e^1$ and $e^4\circ e^3=e^2$ then we may take as
 our symplectic form and metric
 $$[\omega^{ij}] =
 \left(\begin{array}{cccc}0&0&a&b\\0&0&b&c\\-a&-b&0&d-u^2\\-b&-c&-d+u^2&0\end{array}\right)$$
 and
 $$[g^{ij}] =
 \left(\begin{array}{cccc}0&0&0&e\\0&0&-e&0\\0&-e&f&g\\e&0&g&h\end{array}\right)\,,$$
 for any choice of the constants $a, b, c, d, e, f, g, h$
 such that $e\neq 0$ and $b^2\neq ac$.

 This algebra, essentially $(57)_{-1}$, is the only algebra in
 \cite{baimenghe02} of dimension 2 or 4 which admits
 non-degenerate forms $(\cdot,\cdot)$ and
 $\langle\cdot,\cdot\rangle$ satisfying the above compatibility conditions with $\circ$,
 other than the trivial case in which all products are zero, i.e. in which
 the Hamiltonian operators share the same flat connection, and so
 are simultaneously constant.
 \end{example}

 \begin{proposition}\label{prop:atensor}
 If $P_2$ is an infinitesimal deformation of $P_1$ then
 there exists a tensor field $A^i_j$ such that
 \begin{eqnarray}
 a^{ij} &=& g^{ir}A^j_r-g^{jr}A^i_r\,,\nonumber\\
 b^{ij}_k &=&
 2g^{is}A^j_{s,k}-g^{jr}A^i_{k,r}-g^{is}A^j_{k,s}\,,\nonumber\\
 c^{ij}_{kl} &=& g^{is}A^j_{s,kl}-g^{is}A^j_{(k,l)s}\,,\nonumber\\
 c^{ij}_k &=& g^{is}A^j_{s,k}-g^{is}A^j_{k,s}
 \label{eqn:p1p2A}
 \end{eqnarray}
 in flat coordinates for $g^{ij}$. Further, any (1,1)-tensor field
 $A^i_j$ produces an infinitesimal deformation of $P_1$ by the above
 formulae.
 \end{proposition}
 \begin{proof}
 Using the non-degeneracy of $g^{ij}$, we introduce objects
 $\theta^k_{ij}$ and $\phi_{ij}$ by
 \begin{eqnarray*}
 c^{ij}_k &=& g^{ir}\theta^j_{rk}\,,\\
 a^{ij} &=& g^{ir}g^{js}\phi_{rs}\,.
 \end{eqnarray*}
 Then condition \ref{eqn:d12compat:1} of Theorem
 \ref{thm:d12compat} is equivalent to
 $\theta^k_{ij}=-\theta^k_{ji}$, and so we regard $\theta^k_{ij}$
 as a family of 2-forms $\theta^k$ indexed by $k$.

 Condition \ref{eqn:d12compat:3} is equivalent to
 $\theta^k_{jl,i}=\theta^k_{il,j}-\theta^k_{ij,l}$,
 so that $d\theta^k=0$ for each $k$. This allows us to introduce a family
 of 1-forms $\psi^k$ such that
 $$\theta^k_{ij}=(d\psi^k)_{ij}=\psi^k_{i,j}-\psi^k_{j,i}\,.$$
 Each $\psi^k$ can be adjusted by the addition of the exterior
 derivative, $df^k$, of some function $f^k$ without affecting the
 value of $\theta^k_{ij}$.

 Writing $\alpha_{ij}=\phi_{ij}-g_{jr}\psi^r_i+g_{jr}\psi^r_k$,
 we find that condition \ref{eqn:d12compat:4} is equivalent to the
 closedness of the 2-form $\alpha_{ij}$, upon substituting
 $\phi_{ij}$ and $\psi^i_j$ for $a^{ij}$ and $c^{ij}_k$. Thus we
 may introduce a 1-form $h$ with components $h_i$ such that
 $\alpha_{ij}=h_{i,j}-h_{j,i}$, and so
 $$\phi_{ij}=g_{jr}\psi^r_i-g_{jr}\psi^r_j+h_{i,j}-h_{j,i}\,.$$

 If we now let $A^i_j=\psi^i_j+(g^{ir}h_r)_{,j}$ then we have
 $\theta^k_{ij}=A^k_{i,j}-A^k_{j,i}$ and
 $\phi_{ij}=g_{jr}\psi^r_i-g_{ir}\psi^r_j$, so that the two
 equations $a^{ij}=g^{ir}A^j_r-g^{jr}A^i_r$ and
 $c^{ij}_k=g^{ir}A^j_{r,k}-g^{jr}A^j_{k,r}$ are satisfied. The remaining to
 equations follow easily from $c^{ij}_{kl}=c^{ij}_{k,l}$ and
 $b^{ij}_k=a^{ij}_k+c^{ij}_k+c^{ji}_k$.

 For the converse, it is easy to check that conditions
 \ref{eqn:d12compat:1}-\ref{eqn:d12compat:4} of Theorem
 \ref{thm:d12compat} follow from (\ref{eqn:p1p2A}) for any tensor
 field $A^i_j$.
 \end{proof}

 As with Proposition \ref{prop:bvector}, Proposition
 \ref{prop:atensor} may be understood as asserting the existence
 of an evolutionary vector field
 $$e=A^i_j\left(u(x)\right)u^j_x(x)\dbyd{u^i(x)}+\dots$$
 satisfying
 $P_2=-\mathcal{L}_eP_1$
 whenever $P_2$ is an infinitesimal deformation of $P_1$. This is
 therefore not a surprising result; in \cite{getzlerdarboux} Getzler showed
 the triviality of infinitesimal deformations of Hydrodynamic type Poisson brackets. With this, Proposition
 \ref{prop:atensor} can be looked upon as a proof of
 Theorem \ref{thm:d12compat}.

 There is a freedom in $A^i_j$ of $A^i_j\mapsto
 A^i_j+g^{ir}f_{,rj}$ for some function $f$, which does not affect
 the coefficients of $P_2$. This corresponds to adjusting $e$ by a
 Hamiltonian vector field, $e\mapsto e+P_1(\delta f)$.

 If, with reference to Lemma \ref{lem:3equiv}, we impose the
 additional constraint on (\ref{eqn:p1p2A}) that
 $b^{ij}_k=2c^{ij}_k$ then we have the potentiality condition
 $g_{jr}A^r_{k,i}=g_{ir}A^r_{k,j}$, so that there exists a 1-form
 $B_k$ such that
 \begin{equation}\label{eqn:miuraAB}
 A^i_j=g^{ir}B_{j,r}\,.
 \end{equation}
 In this case $a^{ij}=g^{ir}g^{jr}(B_{r,s}-B_{s,r})=g^{ir}g^{jr}(dB)_{rs}$ and the
 freedom $A^i_j\mapsto A^i_j+g^{ir}f_{,rj}$ is $B\mapsto B+df$.
 This means that $B$ can be determined purely from $g^{ij}$ and
 $a^{ij}$, and thus there is no freedom in the choice of
 $c^{ij}_k$ and $c^{ij}_{kl}$. In fact we may write explicitly
 \begin{equation}\label{eqn:uniinfdef}
 c^{ij}_k=g^{js}g_{kr}\dfbyd{a^{ir}}{u^s}\,,\quad
 c^{ij}_{kl}=c^{ij}_{(k,l)}\,,
 \end{equation}
 and with this, $P_2$ is an infinitesimal deformation of $P_1$ if
 and only if
 \begin{equation}\label{eqn:uniinfdefcond}
  g^{ir}a^{jk}_{,r}+g^{jr}a^{ki}_{,r}+g^{kr}a^{ij}_{,r}=0\,.
 \end{equation}

 \begin{corollary}
 Given a flat metric $g$ and a symplectic form $\omega$, there is
 at most one choice of flat symplectic connection $\nabla$ such
 that the degree 2 Hamiltonian operator specified by
 $(\omega,\nabla)$ is compatible with the hydrodynamic operator
 specified by $g$.
 \end{corollary}

 Clearly, if this connection exists it is given
 by(\ref{eqn:uniinfdef}), so this definition must be checked
 against Theorem \ref{thm:hamcond} to verify
 $$P_2^{ij}=\omega^{ij}\left(\frac{d~}{dx}\right)^2
 +2c^{ij}_ku_x^k\frac{d~}{dx} +c^{ij}_{kl}u_x^ku_x^l
 +c^{ij}_ku_{xx}^k$$
 is Hamiltonian. Since equation (\ref{eqn:uniinfdefcond}) is a
 consequence of the antisymmetry of $P_2$, compatibility with the
 Hydrodynamic operator follows immediately.

 We conclude this section with an example of this type.

 \begin{example}
 The Kaup-Broer system \cite{Oevellaxpoisson},
 $$\left(\begin{array}{c}u^1_t\\u^2_t\end{array}\right) =
 \left(\begin{array}{c}u^1_{xx}+2u^2_x+2u^1u^1_x\\-u^2_{xx}+2(u^1u^2)_x\end{array}\right)\,,$$
 is described by the pair of compatible Hamiltonian operators
 \begin{eqnarray*}
 P_1 &=&
 \left(\begin{array}{cc}0&1\\1&0\end{array}\right)\frac{d}{dx}\,,\\
 P_2 &=&
 \left(\begin{array}{cc}0&1\\-1&0\end{array}\right)\left(\frac{d}{dx}\right)^2
 +\left(\begin{array}{cc}2&u^1\\u^1&2u^2\end{array}\right)\frac{d}{dx}
 +\left(\begin{array}{cc}0&u^1_x\\0&u^2_x\end{array}\right)\,.
 \end{eqnarray*}

 Scaling $x\mapsto\varepsilon x$, $t\mapsto\varepsilon t$ splits
 $P_2$ into $P_2^{(1)}+\varepsilon P_2^{(2)}$ where
 \begin{eqnarray*}
 P_2^{(1)} &=&
 \left(\begin{array}{cc}2&u^1\\u^1&2u^2\end{array}\right)\frac{d}{dx}
 +\left(\begin{array}{cc}0&u^1_x\\0&u^2_x\end{array}\right)\,,\\
 P_2^{(2)} &=& \left(\begin{array}{cc}0&1\\-1&0\end{array}\right)
 \left(\frac{d}{dx}\right)^2\,.
 \end{eqnarray*}
 Since $P_2=P_2^{(1)}+\varepsilon P_2^{(2)}$ is Hamiltonian for
 all $\varepsilon$, $P_2^{(1)}$ and $P_2^{(2)}$ constitute a bi-Hamiltonian structure of
 the type considered above. A set of flat coordinates for the metric in
 $P_2^{(1)}$ is
 \begin{eqnarray*}
 {\tilde u}^1 &=& u^1 \,,\\
 {\tilde u}^2 &=& \sqrt{4u^2-(u^1)^2}\,,
 \end{eqnarray*}
 in which
 \begin{eqnarray*}
 {\tilde P}_2^{(1)} &=&
 \left(\begin{array}{cc}2&0\\0&2\end{array}\right)\frac{d}{dx}\,,\\
 {\tilde P}_2^{(2)} &=& \frac{2}{{\tilde
 u}^2}\left(\begin{array}{cc}0&1\\-1&0\end{array}\right)\left(\frac{d}{dx}\right)^2
 +\frac{4}{({\tilde u}^2)^2}
 \left(\begin{array}{cc}0&-{\tilde u}^2_x\\0&{\tilde
 u}^1_x\end{array}\right)\frac{d}{dx}\\
 && +\frac{4}{({\tilde u}^2)^3}
 \left(\begin{array}{cc}0&({\tilde u}^2_x)^2\\0&-{\tilde u}^1_x{\tilde u}^2_x\end{array}\right)
 +\frac{2}{({\tilde u}^2)^2} \left(\begin{array}{cc}0&-{\tilde u}^2_{xx}\\0&{\tilde
 u}^1_{xx}\end{array}\right)\,.
 \end{eqnarray*}

 So in this situation we have, for the 1-form in
 (\ref{eqn:miuraAB}),
 $$B=\frac{{\tilde u}^1}{2{\tilde u}^2}d{\tilde u}^2\,.$$
 \end{example}

 \section{Conclusions}
 In section \ref{section:pencils} an approach was taken based upon
 the methods of \cite{dubflatpencil} to study compatible pairs of
 Hamiltonian operators of degree 2 which satisfy the conditions of
 the relevant Darboux theorem, Theorem \ref{thm:H2ODarboux}. As
 for Hydrodynamic Poisson pencils, the compatibility could be
 reduced to algebraic constraints on a multiplication of
 covectors. Driving this was the ability to reduce a given
 Hamiltonian operator on $L(M)$ to a flat Fedosov structure
 $(\omega,\nabla)$ on $M$, which are natural symplectic analogues
 of the pair consisting of a flat metric and its Levi-Civita
 connection which determines a Hydrodynamic Poisson bracket.

 To extend such a results to pairs of arbitrary degree 2 Hamiltonian
 operators, one must consider the pair $(a,\nabla)$ of Theorem
 \ref{thm:adela}. The condition (\ref{eqn:bivectcompat}), whilst
 atypical, expresses a familiar concept; in almost-symplectic
 geometry, it is common to consider connections such that the
 covariant derivative of the almost-symplectic form is zero, but
 which have torsion; if the torsion of such a connection is
 skew-symmetric then its symmetric part satisfies
 (\ref{eqn:bivectcompat}). Equation (\ref{eqn:covbfromc}) provides
 the means of going from the symmetric connection to the compatible
 connection with skew-torsion. The only formula missing above
 necessary to the study of arbitrary bi-Hamiltonian structures
 of degree 2 is an expression for the contravariant curvature of
 the connection defined by $c^{ij}_k$, which is,  in the presence
 of Theorem \ref{thm:hamcond}'s condition \ref{eqn:hamcond:deriv},
 $$R^{ijk}_l=a^{ir}(c^{jk}_{r,l}-c^{jk}_{l,r}) +c^{ij}_rc^{rk}_l
 +c^{ik}_rc^{rj}_l -(b^{ij}_r-2c^{ij}_r)c^{rk}_l
 +c^{ik}_r(b^{rj}_l-2c^{rj}_l)\,.$$
 One may use \ref{eqn:hamcond:deriv} to replace the components of $b^{ij}_k$
 in this expression with those of $c^{ij}_k$ and the derivatives of
 $a^{ij}$. However, one sees that the compatibility conditions do
 not naturally become algebraic constraints on $\Delta^{ij}_k$,
 and the relevancy of such an approach is undermined. It is
 interesting to note, however, that equation (\ref{eqn:c4c}) still
 holds (with $\Gamma^{ij}_k=c^{ij}_k$), so that $\circ$ defined by
 $\Delta^{ij}_k$ still satisfies $(I\circ J)\circ K=(I\circ
 K)\circ J$, and that it is the {\lq Fermionic\rq} condition
 $(I\circ J)\circ K=-(I\circ K)\circ J$ which is altered.

 The proof of Proposition \ref{prop:bvector} is easily adapted to
 confirm the existence of a vector field $B$ realising
 $P_1=-\mathcal{L}_BP_2$ whenever $P_1$, of the form
 (\ref{def:h20}) is an infinitesimal deformation of $P_2$ as a
 Hamiltonian operator, provided ${b_1}^{ij}_k=2{c_1}^{ij}_k$. A
 simple calculation of $\mathcal{L}_BP_2$ for arbitrary $B$ shows
 that ${b_1}^{ij}_k=2{c_1}^{ij}_k$ is also a necessary condition.
 Thus we have determined the trivial deformations of a degree 2
 Hamiltonian operator admitting a constant form, which are
 themselves of degree 2. Clearly a different approach is necessary
 to understand deformations of higher degrees. For the case of
 operators not satisfying the constraints of Theorem
 \ref{thm:H2ODarboux}, it is not immediately obvious what
 conditions, if any, will guarantee the triviality of a
 deformation; owing to the different form the contravariant
 curvature tensor takes, the condition
 ${c_1}^{ij}_{k,l}={c_1}^{ij}_{l,k}$ is absent. Owing to the lack
 of a constant form, the methods of \cite{dubzhangnorm} in
 ascertaining the triviality of higher degree deformations, if
 applicable, will be somewhat more complicated.

 Finally, there is a certain artificiality to the examples of
 compatible Fedosov structures presented in section
 \ref{section:pencils}. Given Theorem \ref{thm:fdbh}'s assertion
 that underlying a pair of compatible Fedosov structures is a
 finite-dimensional bi-Hamiltonian structure, the question is
 raised asking which finite-dimensional bi-Hamiltonian structures
 admit symplectic connections forming almost compatible, almost
 compatible and flat, or compatible Fedosov structures? It would
 be interesting to exhibit a pair of compatible Fedosov structures
 in which the flat coordinates for one of the connections are in
 some sense physical.

 \section*{Acknowledgements}
 The author would like to thank Ian Strachan for suggesting this
 project, and the Carnegie Trust for the Universities of Scotland
 for the scholarship under which this work was conducted.

\bibliographystyle{plain}

\end{document}